\documentclass[journal]{IEEEtran}
\ifCLASSINFOpdf
\else
\fi
\usepackage{cite}
\usepackage[cmex10]{amsmath}
\usepackage{amssymb}
\usepackage{algorithmic}
\usepackage{array}
\usepackage{mdwmath}
\usepackage{booktabs}
\usepackage{mdwtab}
\usepackage{eqparbox}
\usepackage{graphicx}
\usepackage[justification=centering]{caption}
\usepackage{color}
\usepackage{bm}
\usepackage{float}
\begin{document}

\title{ Identifying the Mislabeled Training Samples of ECG Signals using Machine Learning }

\author{Yaoguang~Li, ~Wei~Cui~\IEEEmembership{Senior Member,~IEEE}£¬ and Cong Wang~\IEEEmembership{Senior Member,~IEEE}

\thanks{
This work is supported by the National Natural Science Foundation of China under Grant 11404113, and the Guangzhou Key Laboratory of Brain Computer Interaction and Applications under Grant 201509010006. \textit{Corresponding
author: Wei Cui.}

The authors are with the School of Automation Science and Engineering, South China University of Technology, Guangzhou 510641, China~(e-mails: 852934033@qq.com, aucuiwei@scut.edu.cn, wangcong@scut.edu.cn).
}
}

\markboth{Submitted to IEEE Journal of Biomedical and Health Informatics}%
{Shell \MakeLowercase{\textit{et al.}}: Bare Demo of IEEEtran.cls for Journals}
\maketitle

\begin{abstract}
 The classification accuracy of electrocardiogram signal is often affected by diverse factors in which mislabeled training samples issue is one of the most influential problems. In order to mitigate this negative effect, the method of cross validation is introduced to identify the mislabeled samples. The method utilizes the cooperative advantages of different classifiers to act as a filter for the training samples. The filter removes the mislabeled training samples and retains the correctly labeled ones with the help of 10-fold cross validation. Consequently, a new training set is provided to the final classifiers to acquire higher classification accuracies. Finally, we numerically show the effectiveness of the proposed method with the MIT-BIH arrhythmia database.
\end{abstract}

\begin{IEEEkeywords}
ECG signal, mislabeled samples, cross validation, machine learning.
\end{IEEEkeywords}

\IEEEpeerreviewmaketitle

\section{Introduction}
\IEEEPARstart{C}{ardiac} disease has become a key issue that threatens human life safety. How to detect heart disease as early as possible is the core of this problem, because cardiac disease is very prone to sudden death, which makes the detection of heart disease more urgent than treatment.
 Electrocardiogram~(ECG) signal contains a lot of useful information that can be used to diagnose various cardiac diseases. Nowadays, ECG is the most effective tool for heart disease detection and the use of machine learning algorithms for automatic detection of ECG signal has become an increasingly significant topic in the relevant areas~\cite{Batra:2016,Rannal:2016,Huanhuan:2014,Exarchos:2006}. Moreover, body sensor network-based devices have become widely accepted and novel ECG telemetry systems for cardiac health monitoring applications have been proposed \cite{Lee:2015,Chen:2015,Wang:2010}.
 In order to improve the classification accuracy of ECG signal, most works focus on two aspects,~(1) feature selection~\cite{Chazal:2004,Inan:2006}, and~(2) robustness of the machine learning classifiers~\cite{Inan:2006,Ince:2008}. It is clear that the use of ECG signal features that maximize the distinction between different diseases can significantly improve the classification accuracy such as temporal intervals~\cite{Chazal:2004,Inan:2006,Ince:2008}, morphological features~\cite{Chazal:2004}, frequency domain features \cite{Lin:2008}, high-order statistics~\cite{Osowski:2004}, wavelet transform coefficients~\cite{Inan:2006,Ince:2008,Daamouche:2012}. There are also some studies on discriminant function optimization of different classification algorithms~\cite{Inan:2006,Fayn:2011}. Ref.\cite{Chazal:2004} shows that linear models achieve good classification accuracies, nevertheless, more interests have been attracted to the nonlinear approaches in the past few years. Neural network~\cite{Ince:2008,Lin:2008,Martis:2013,Jiang:2007} and support vector machine~(SVM)~\cite{Osowski:2004,Daamouche:2012,Kampouraki:2009} are the most popular algorithms. In addition, some optimization algorithms such as genetic algorithm~\cite{Nasiri:2009} and particle swarm algorithm~\cite{Melgani:2008,Korurek:2010} are used to optimize the parameters in the classification algorithms to improve the classification accuracy.

\begin{figure}
  \centering
  \includegraphics[width=8cm]{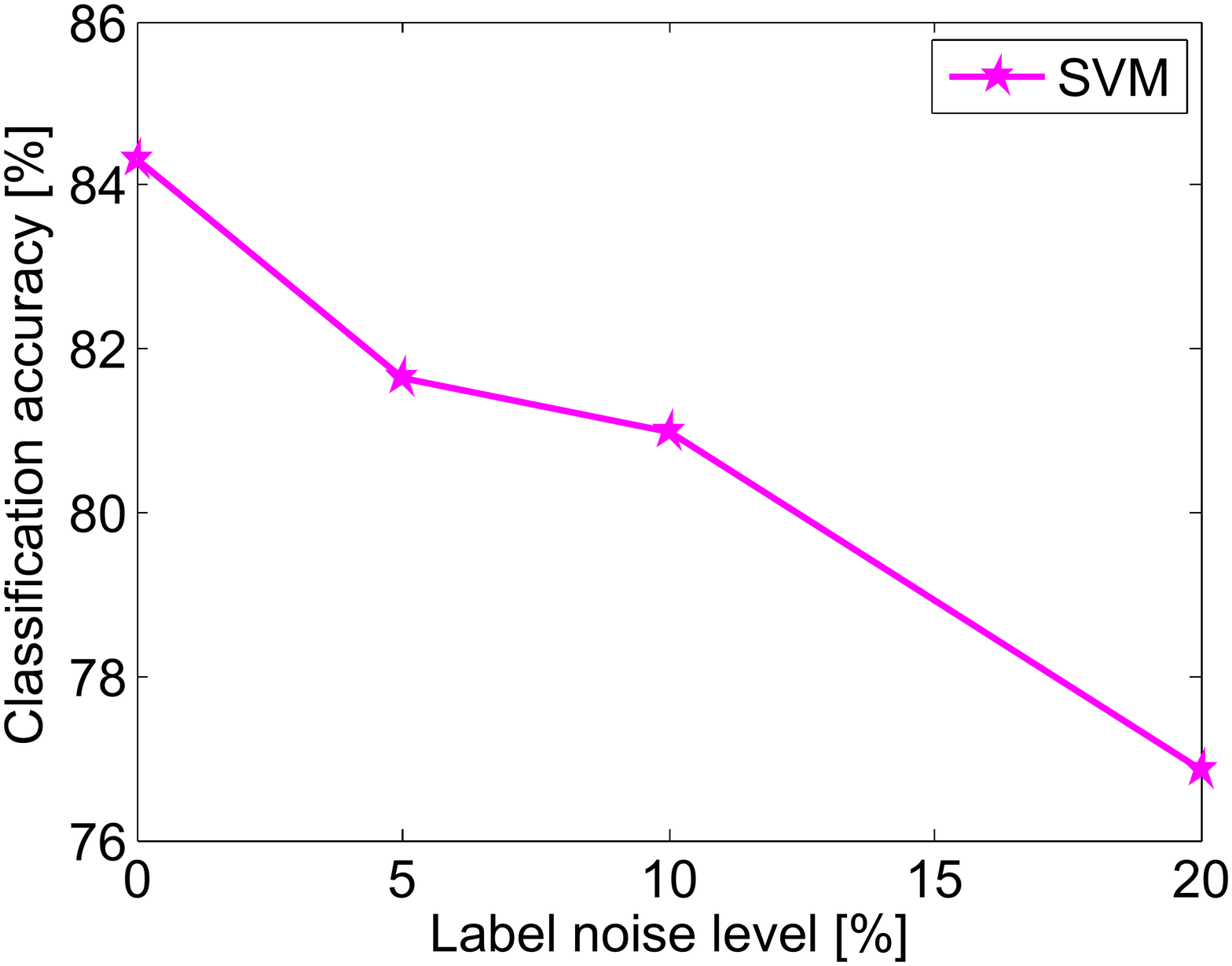}\\
  \caption{Classification accuracy with label noise in Ref.\cite{Pasolli:2015}}
  \label{fmzhexiantu}
\end{figure}

\begin{figure}
  \centering
  \includegraphics[width=8cm]{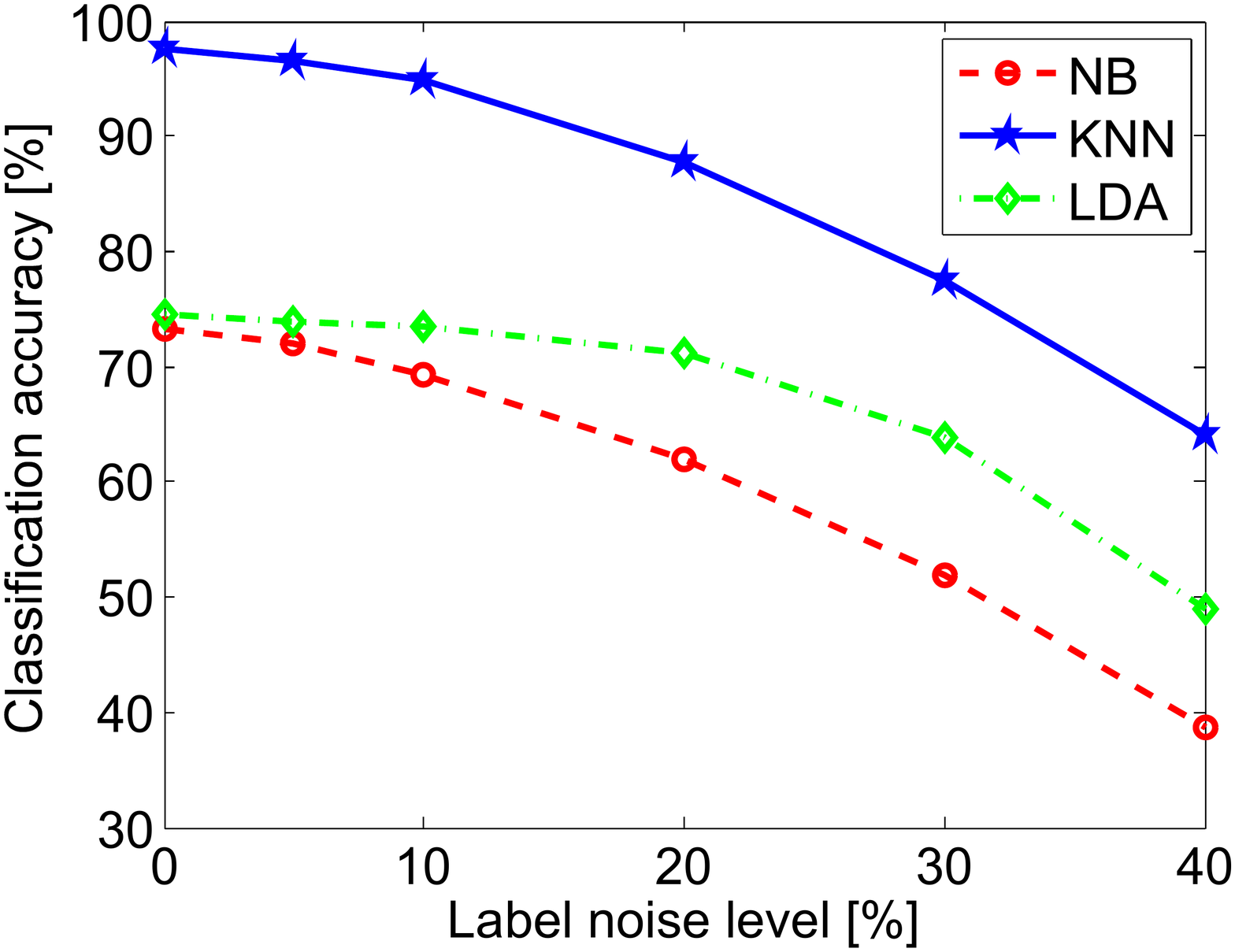}\\
  \caption{Classification accuracy with label noise in this paper}
  \label{benzhexiantu}
\end{figure}

Although these methods mentioned above have been proved to be effective in previous works, they are based on an essential assumption that the samples used to train the classifier are completely reliable, which can not always be guaranteed in the real world. Many factors may make ECG data less reliable such as medical expert diagnosis error, data encoding and communication problems. The unreliable issues are distinguished into two types: feature~(or attribute) noise and class~(label) noise. In particular, if the training set has mislabeled samples, the classifier will be deteriorated heavily, and remarkably reduce the actual classification accuracy. Therefore, the automatic analysis of ECG signal with computer technology is still an auxiliary equipment in the cardiac disease diagnosis. {\color{red}In} Ref.\cite{Brodley:1999}, the mislabeled samples issue is the most prominent factor that reduces the classification accuracy of ECG signal. In Ref.\cite{Frenay:2014}, it is summarized that with mislabeled samples, the prediction accuracy decreases quickly and the complexity of classifiers increases rapidly. For instance, Ref.\cite{Frenay:2014} and Ref.\cite{Quinlan:1986} demonstrate that the model of decision tree~(DT) would become more complicated with label noises. In Fig.~\ref{fmzhexiantu} and  Fig.~\ref{benzhexiantu}, we show the classification accuracies with different label noises levels in Ref.\cite{Pasolli:2015} and this manuscript, respectively. Obviously, even the four commonly used machine learning algorithms are adopted, the classification accuracies of ECG signal are still deteriorated badly due to the presence of mislabeled samples in the training set.

Ref.\cite{Louis:2016} takes advantage of the Gaussian mixture model to model normal ECG heartbeat and develops a real-time abnormal ECG heartbeat detection and removal system. Ref.\cite{Pasolli:2015} proposes an optimal subset search algorithm based on the genetic algorithm to achieve the highest classification accuracy in the test set. It is considered that the training samples outside of the optimal subset are regarded as the mislabeled, which need to be removed from the original training set. Even if some of the mislabeled samples are identified correctly by this method, the identification rate is not high enough and the classification result is still unsatisfactory. Besides, the highest label noise level in the original training set considered in Ref.\cite{Pasolli:2015} is only 20\%.

The purpose of this manuscript is to enhance the identification rate of the mislabeled samples and improve the classification accuracy of ECG signal with the cross validation method. Compared with the previous studies, the proposed method can achieve higher identification rate of mislabeled samples. Moreover, if the mislabeled samples existed in the training set is less than 20\%, the classification accuracy can be increased to the same level as there is no mislabeled sample in the training set with the help of the proposed method. If the proportion of mislabeled samples is 30\%, the classification accuracy is slightly lower than the case of no mislabeled sample in the training set. If the proportion meets 40\%, the classification accuracy is still much higher than the circumstance without filtering.

The rest of the paper is organized as follows. Section 2 introduces the cross validation method and machine learning algorithms briefly. In Section 3, We summarize the basic procedure of ECG signal processing, including data collection, data preprocessing and {feature extraction} etc. Section 4 presents the experiment results with the MIT-BIH arrhythmia database {\cite{Mark:1997}}. Finally, We conclude our results in Section 5.

\section{Cross validation and Machine learning algorithms}
\subsection{Cross validation}
Cross validation~\cite{Arlot:2010,Zhang:2015} is a statistical analysis method used to verify the performance of a classifier. Its basic idea is to divide the original data into two parts. One part serves as the training set, and the other serves as the validation set. The training set is used to train the classifier, and then the classifier is tested on the validation set to evaluate its performance. There are four main cross validation methods.

a.~Hold-out method:~The original data is randomly divided into two groups, one as the training set, and the other as the validation set. The training set is used to train the classifier, and then the classifier is tested on the validation set to evaluate its performance of classification accuracy. Because the way to divide the original data into groups by the hold-out method is random, the final classification accuracy on the validation set has a direct relationship to the result of grouping, which makes the performance evaluation of the hold-out method not convincing enough.

b.~Double cross validation~(2-CV):~2-CV divides the data into two equal-sized subsets and performs two rounds of classifier training. In the first round, a subset serves as the training set and the other is used for the validation set. In the second round, the training set and the validation set are interchanged to train the classifier again. The two classification accuracies represent the performance of the classifier. But the 2-CV is not commonly used, because the number of samples in the training set is so small that the training set can not represent the distribution of all the data.

c.~K-fold cross validation~(K-CV):~The original data is divided into $k$ groups~(usually equalized). Each group serves as the validation set for one time with the remaining $k-1$ groups as the training set. Then we obtain $k$ models and the corresponding $k$ classification accuracies as the performance of the classifier. The number $k$ is usually larger than three. The K-CV can effectively avoid over-learning and under-learning and obtain persuasive results.

d.~Leave-one-out cross validation~(LOO-CV):~If there are $N$ samples in the original data, then the LOO-CV is the same as N-CV, which means each sample serves as the validation set and the rest $N-1$ samples are the training set. Then we obtain $N$ models and the average of the classification accuracies of the $N$ models serves as the performance of the classifier. Compared with the K-CV and the 2-CV, in the LOO-CV, almost all the samples of the original data are used to train the model. So the training set of the LOO-CV can represent the distribution of all the original data and the results of the performance evaluation are more reliable. Moreover, there is no random factor to make the experiment unrepeatable in the LOO-CV. But the high computational cost makes the LOO-CV unworkable when the data size is large.

In this paper, 10-CV is adopted in view of the maximization of the use of the original training samples and the minimization of computational consumption.

\subsection{Machine learning algorithm}
\subsubsection{Support vector machine~(SVM)}

SVM~{\cite{Theodoridis:2008,Cristianini:2000,Suykens:2001}} maps the feature vector $x\in{R^d}$ to the high-dimensional feature space $\phi(x)\in H$, and creates an optimal separation hyperplane that maximizes the interval between support vectors and the hyperplane. Different SVM classifiers are generated by different mapping rules. The mapping function $\phi(\cdot)$ is determined by the kernel function $ K(x_i,x_j)$, which defines an inner product of $ H $ space.

The optimization problem of the SVM interval can be written as follows:
\begin{eqnarray}
&&\left\{
    \begin{aligned}
    &\max\limits_{\alpha} ~~  \sum_{i=1}^N {\alpha_i} - \frac{1}{2}\sum_{i=1}^N \sum_{j=1}^N \alpha_i\alpha_j\cdot y_i y_j \cdot K(x_i,x_j)\\
    &~\text{s.t.}  ~~~~\sum_{i=1}^N{\alpha_i y_i}=0\\
    &~~~~~~~~~~~0\leqslant\alpha_i\leqslant C~~~~~i=1,2,\dots,N,
    \end{aligned}
\right.
\end{eqnarray}
and the SVM decision function is
\begin{eqnarray}
f(x)=\text{sgn}\left(\sum_{i=1}^N y_i\alpha_i\cdot K(x,x_i)+b \right).
\end{eqnarray}

\subsubsection{K-nearest neighbor~(KNN)}

In KNN classification~{\color{red}\cite{Coomans:1982,Cover:1967,Zhang:2007,Keller:1985}}, the output is a class membership. An object is classified by a majority vote of its $k$ nearest neighbors~($k$ is a positive integer, typically small). If $k = 1$, the object is simply assigned to the class of its single nearest neighbor.

\subsubsection{Naive Bayes~(NB)}
The bayes classification method~\cite{Viaene:2004,Kim:2006} classifies a certain sample based on its posterior probability calculated by priori probability and data.

In bayes classifier, if D is a certain sample with the feature $x=(x_1,x_2,\dots,x_n )$, its probability of belonging to label $y_i$ is: $P(Y=y_i|X_1=x_1,X_2=x_2,\dots,X_n=x_n),(i=1,2,\dots,m)$, that is
\begin{equation}
\begin{aligned}
P(Y&=y_j|X=x)=\max\{ P(Y=y_1|X=x),\\ & P(Y=y_2|X=x), \dots, P(Y=y_m|X=x)\}.
\end{aligned}
\end{equation}
Based on the bayesian formula:
\begin{eqnarray}
P(Y=y_j|X=x)=\frac{P(X=x|Y=y_j)P(Y=y_j)}{P(X=x)},
\end{eqnarray}
the key point to obtain the posterior probability is to calculate $P(X=x|Y=y_j)$.

It is difficult to calculate $P(X=x|Y=y_j)$ directly. So NB is created with the assumption that the features of the object are independent of each other, that is $P(X=x|Y=y_j)=P(x_1|y_j)P(x_2|y_j)\cdots P(x_n|y_j)$. Consequently, the posterior probability of each class can be calculated and the class with the largest posterior probability is just the classification label of the sample.
\subsubsection{Linear discriminant analysis~(LDA)}
The principle of the LDA~{\cite{Fisher:1936,McLachlan:2004,Ye:2004,Haeb-Umbach:1992}} is that the data points~(vectors) are projected to the space of lower dimension so that the projected vectors can be easily classified. Thus, the points that belong to the same type are much closer than those belong to different types in the projection space. Suppose that the projection function is $y=w^{T}x$, we can calculate

a. ~the original center point~(mean value) of the class $i$ is
\begin{equation}
 m_i=\frac{i}{n_i}\sum_{x\in D_i}x,
\end{equation}
where $D_i$ refers to the set of points that belong to class $i$.

b. ~The center of class $i$ after projection is
\begin{equation}
\widetilde{m_i}=w^{T}m_i.
\end{equation}

c. ~The degree of dispersion (variance) between classes is
\begin{equation}
\widetilde{S_i}=\sum_{y\in{Y_i}}{(y-\widetilde{m_i})^2}.
\end{equation}

d. ~The objective optimization function of LDA after projection to $w$ is
\begin{eqnarray}
J(w)=\frac{|\widetilde{m_1}-\widetilde{m_2}|^2}{{\widetilde{S_1}}^2+\widetilde{S_2}^2}.
\end{eqnarray}

\subsubsection{Decision tree~(DT)}
A DT is a flowchart-like structure in which each internal node represents a ``test" on a feature, each branch represents the outcome of the test, and each leaf node represents a class label (decision taken after computing all attributes). The paths from root to leaf represent classification rules.

C4.5 is an algorithm used to generate a DT~{\cite{Quinlan:1993,Quinlan:1996,Ruggieri:2002}}. C4.5 is an extension of the earlier ID3 algorithm~\cite{Pal:2001}. The DTs generated by C4.5 can be used for classification. Both C4.5 and ID3 use the information entropy to build DTs from a set of training data. C4.5 makes some improvements to ID3. Some of these are handling both continuous and discrete attributes, handling training data with missing attribute values, handling attributes with differing costs, and pruning trees after creation.

\section{ECG signal processing}
ECG is a technique for recording the changes in electrical activity of each cardiac cycle of the heart, which provides diagnostic information about the cardiac condition of a patient.

\begin{figure}
  \centering
  \includegraphics[width=6cm]{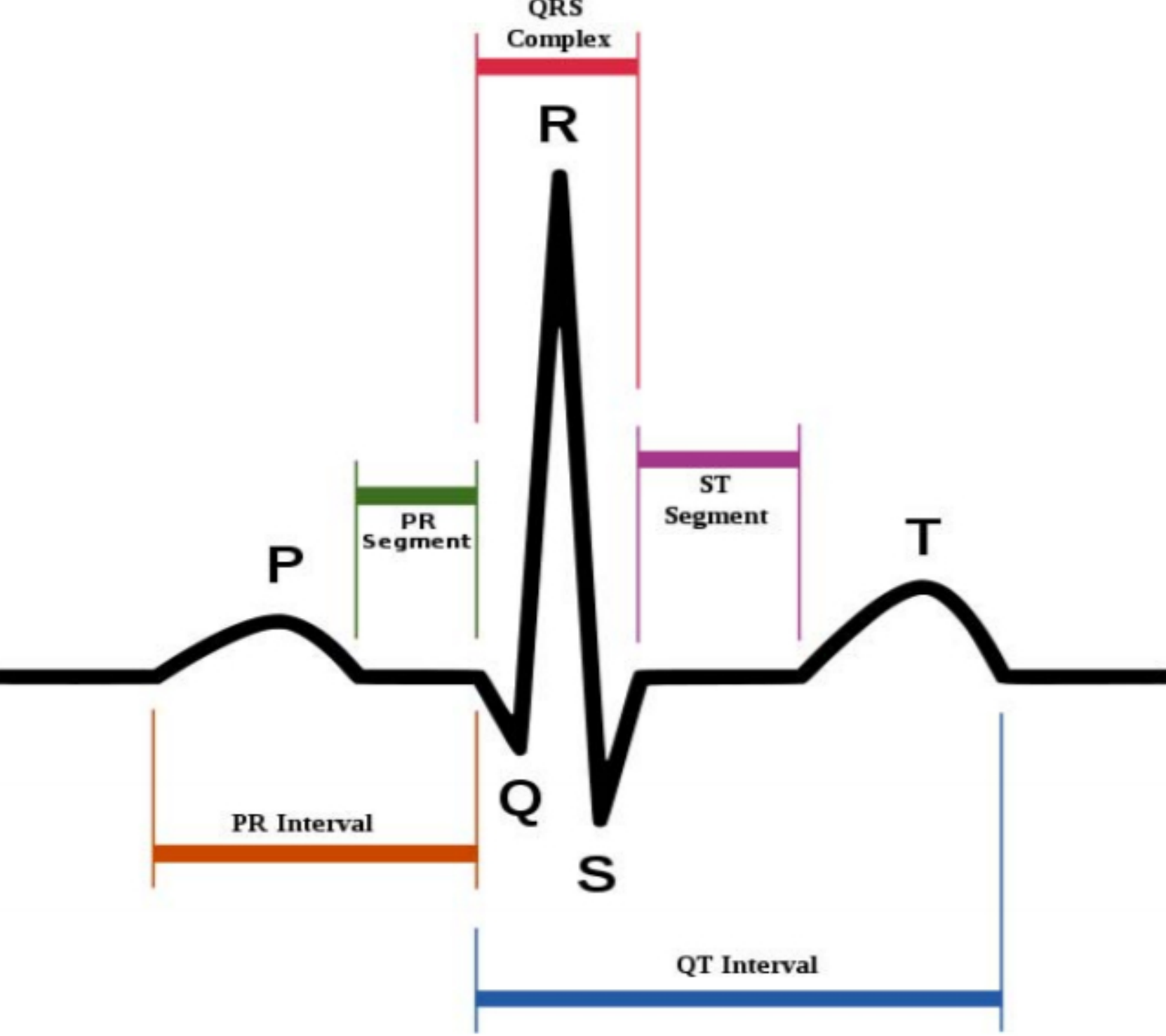}\\
  \caption{A real time ECG signal~\cite{electrocardiograph:2011}}
  \label{xindiantu}
\end{figure}

As shown in Fig.~\ref{xindiantu}, in a normal ECG cycle, P wave is the first upward deflection and represents the electric potential variation of the two atrial depolarization. It has positive polarity and its duration is between 80~ms and 110~ms. The QRS complex wave, with the duration of 60~ms--100~ms, is the most important part in the ECG signal automatic analysis, and represents the electric potential variation of the two ventricular depolarization. Q wave is the first negative waveform, R wave is always the first positive deflection that follows the Q wave, and S wave is the downward deflection after the R wave~\cite{Tabassum:2016}. T wave represents the electric potential variation of the two ventricular repolarization, and its direction is the same as the QRS main wave. We can detect various cardiac abnormalities with the help of the difference in amplitude and duration of different waves. The most distinguishable features of ECG signal are P wave period and peak, QRS-complex period, R-R interval, R peak, etc. Doctors observe the deviations of P, QRS and T waves from the normal signal in terms of interval and amplitude to find the abnormalities in heart.

 Since the computer performs faster and more accurate than the human eyes in observing these features. Thus, it is obvious that the use of computers for ECG signal analysis is a better way in cardiac detection and monitoring. Normal sinus rhythm is the cardiac rhythm that begins with the myocardial contraction of the sinus node. A normal sinus rhythm shows the following five characteristics~\cite{Tabassum:2016}.

a. ~normal P wave,

b. ~constant P-P and R-R interval,

c. ~constant P wave configuration in a given lead,

d. ~P-R interval and QRS interval within normal limit,

e. ~heart rate between 60 to 100 beats/min.

 It is divided into 11 different arrhythmias types in the MIT-BIH arrhythmia database based on the different situations of abnormality or disturbance~\cite{Qin:2015}: left bundle branch block beat, right bundle branch block beat, aberrated atrial premature beat, premature ventricular contraction, fusion of ventricular and normal beat, nodal premature beat, atrial premature beat, premature or ectopic supraventricular beat, ventricular escape beat, nodal escape beat ,and paced beat. There are 12 categories in total with the normal beat.

\begin{figure}
  \centering
  \includegraphics[width=6cm]{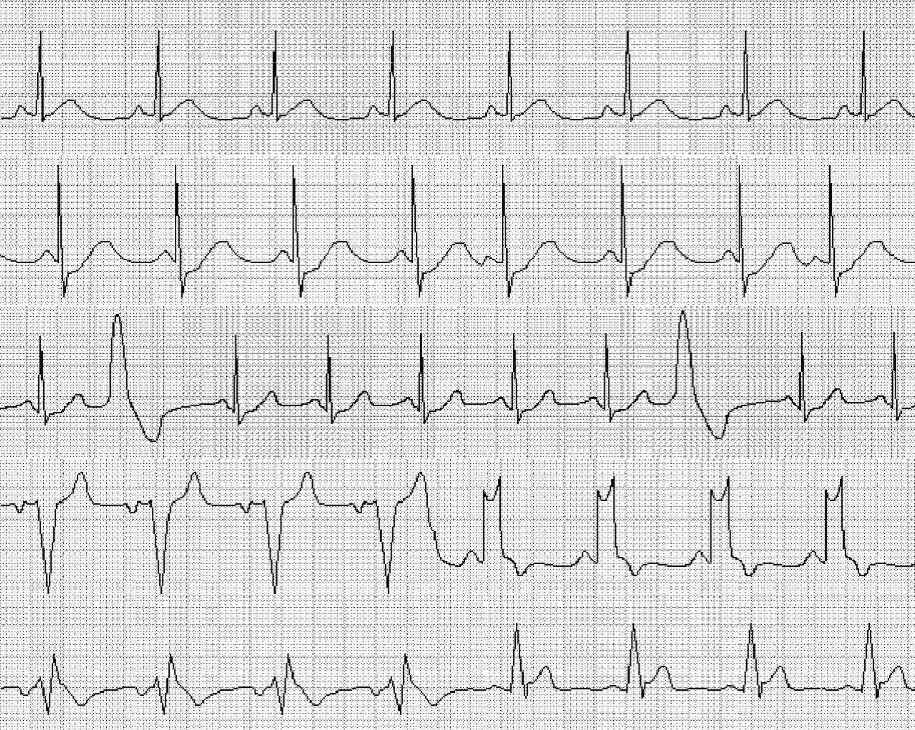}\\
  \caption{{Several common arrhythmia ECG waveforms}}
  \label{arrhythmia}
\end{figure}

 { Fig.~\ref{arrhythmia} describes several common arrhythmia ECG waveforms. From top to bottom are normal beat, atrial premature beat, premature ventricular contraction, left bundle branch block beat and right bundle branch block beat, respectively. Atrial premature beats are characterized by an abnormally shaped P wave. Since the premature beat initiates outside the sinoatrial node, the associated P wave appears different from those seen in normal sinus rhythm. Typically, the atrial impulse propagates normally through the atrioventricular node and into the cardiac ventricles, resulting in a normal, narrow QRS complex. In general, premature ventricular contraction has early appearance of QRS complex, with no previous P wave. The QRS duration is more than 0.12 seconds, with large wave deformity. The ST segment and T wave have opposite direction to QRS main wave with a complete compensation intermittent. Bundle branch block beats' QRS duration is more than 0.12 seconds with possible double R waves (R-r'). Sometimes, we can only find a notch gap between R and r' in the left bundle branch block beat in lead V5, V6. However, the QRS wave of the right bundle branch block beat is like a 'M' shape. Many ECG waveforms of different arrhythmias have similarities, which leads to the possibility of mislabeling especially with the presence of noise.}

\subsection{ECG signal preprocessing}
 The amplitude of a normal ECG signal ranges from $10\mu V$ to $4mV$, thus the ECG signal acquisition and processing should be considered as weak signal detection issues. Based on the properties of ECG signal, an important step is to filter out the general noises in the signal and to retain the valuable components of each waveform. Two main noises in the ECG signal are baseline wander and power interference.

 Baseline wander has the strongest negative impact on ECG signal. It is a low-frequency signal ranging from 0.05Hz to several Hz. We filter out the baseline wander effectively using the median filtering~\cite{Wischermann:1991} in this paper. The formula of the median filtering is
\begin{eqnarray}
 y(i)=\text{Med}\bigg\{x(i-N),\cdots,x(i),\cdots,x(i+N)\bigg\},
\end{eqnarray}
where the $x(i)$ is the original value, $y(i)$ is the new one, and $N$ is adjustable.

Power interference mainly refers to 50Hz and its high harmonic interference. The human body has antenna effects because of its natural physiological characteristics, nevertheless the ECG signal collection equipment usually has long wires exposed with antenna effects, which makes power interference the most common noise in human ECG signal~\cite{Alste:1985}. Due to its good time-frequency localization characteristics, wavelet transform has become a popular method in signal denoising. It decomposes the noisy signal into multi-scale, and then the wavelet coefficients that belong to the power interference are removed according to the frequency ranges of different kinds of signals. Ultimately, the remaining wavelet coefficients are utilized to reconstruct the signal.
\subsection{ECG signal { feature extraction}}
As described in the introduction part, there are different representation types of ECG signal such as temporal intervals, frequency domain features, high-order statistics, and so forth. In order to compare our results with the previous work~\cite{Pasolli:2015}, we adopt the same features~(1) ECG morphology features, and (2) three ECG temporal features, i.e., the QRS complex duration, the RR interval (the time span between two consecutive R points representing the distance between the QRS peaks of the present and previous beats), and the RR interval averaged over the ten last beats. { The morphology features are extracted from the segmented ECG cycles. We define the middle sample of two R-peak samples as M, and a ECG cycle ranges from the previous M of the current R-peak to the next M of the current R-peak. The morphology features are acquired by normalizing the duration of the segmented ECG cycles to the same periodic length according to the procedure reported in~\cite{Wei:2001}}. That is, one of the ECG segments \bm{$y_i$}=$[y_i(1),y_i(2),\cdots,y_i(n^*)] $ can be converted into a segment \bm{$x_i$}=$[x_i(1),x_i(2),\cdots,x_i(n)] $ that holds the same signal morphology, but in different data length (i.e., $n^*\not=n $) using the following equation,
\begin{eqnarray}
 x_i(j)=y_i(j^*)+(y_i(j^*+1)-y_i(j^*))(r_j-j^*),
\end{eqnarray}
 where $r_j=(j-1)(n^*-1)/(n-1)+1 $, and $ j^*$ is the integral part of $r_j $. In this paper, we set $n=300$. Therefore, the various lengths of the ECG segments will be compressed or extended into a set of ECG segments with the same periodic length. Consequently, the total number of morphology and temporal features is equal to 303.

\subsection{Feature normalization and dimensionality reduction}
Temporal features and morphology features have different dimensions, which damages the classification results seriously. Data normalization means adjusting values measured on different scales to a common scale. In this paper, we use the min-max normalization rule~\cite{Jain:2011} to map the original data to 0--1 by linear transformation
\begin{eqnarray}
x^* = \frac{x-min}{max-min},
\end{eqnarray}
where $max$ and $min$ are the maximum and minimum values of the feature dimensions of $x$, respectively.

Finally, due to the high-dimension properties of the features, we use the principal component analysis~(PCA) technique{ \cite{Wold:1987,Jolliffe:2002,Martis:2013}} to project the features into a lower dimensional feature space.

\section{Experiment}
\subsection{Data description}

The method proposed for the mislabeled samples identification is tested experimentally on real ECG signals that obtained from the well-known MIT-BIH arrhythmia database. In particular, in order to facilitate the comparison with Ref.\cite{Pasolli:2015}, the considered beats referred to the following six classes: normal sinus rhythm~(N), atrial premature beat~(A), ventricular premature beat~(V), right bundle branch block~(RB), paced beat~(P), and left bundle branch block~(LB). { According to Ref.~\cite{Pasolli:2015}, the beats were selected from the recordings of 20 patients, which corresponded to the following files:} 100, 102, 104, 105, 106, 107, 118, 119, 200, 201, 202, 203, 205, 208, 209, 212, 213, 214, 215, and 217. With 36328 heart beats in total, there are 24150 N class, 338 A class, 2900 V class, 3689 RB class, 3450 P class and 1801 LB class. { All beats are divided into the training set and the test set by the class distribution.} There are 1500 N class, 100 A class, 1000 V class, 1000 RB class, 1000 P class, 500 LB class, totally 5100 beats in the training set. The rest of the beats are assigned to the test set.

After the data preprocessing, { feature extraction}, normalization and reduction, we establish our experiment in the next subsection.

\subsection{Experiment setup}
{ We assume that the original ECG recording labels of the MIT-BIH arrhythmia database are totally correct.} We add some label noises with different levels~(5\%, 10\%, 20\%, 30\%, 40\%) to the training set, namely, changing the label of some training samples artificially. { However, the test set remains unchanged. That means all the operations are on the training set, and the test set is only used to test the effectiveness of the operations on the training set.} A noise-free training set is used for comparison. The number of the changed labels of each class is based on its proportion to the overall training set. { For example, when label noise is 5\%, the number of changed labels of class N equals 1500*5\%=75.}

\begin{figure*}
  \centering
  \includegraphics[width=13cm,height=5cm]{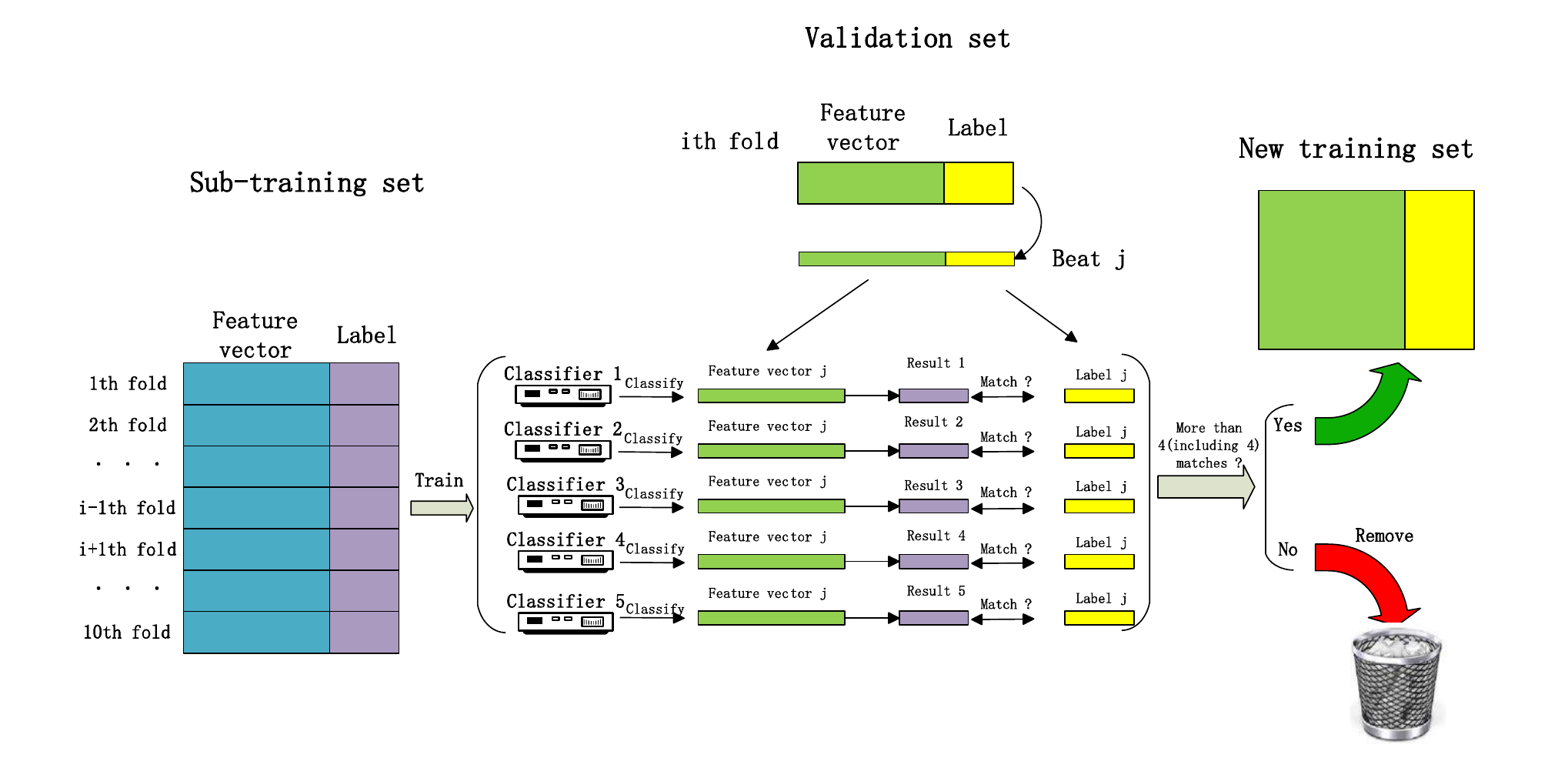}\\
  \caption{Illustration of the proposed automatic training sample validation framework}
  \label{tuhua}
\end{figure*}

\begin{figure}
  \centering
  \includegraphics[width=4.9cm]{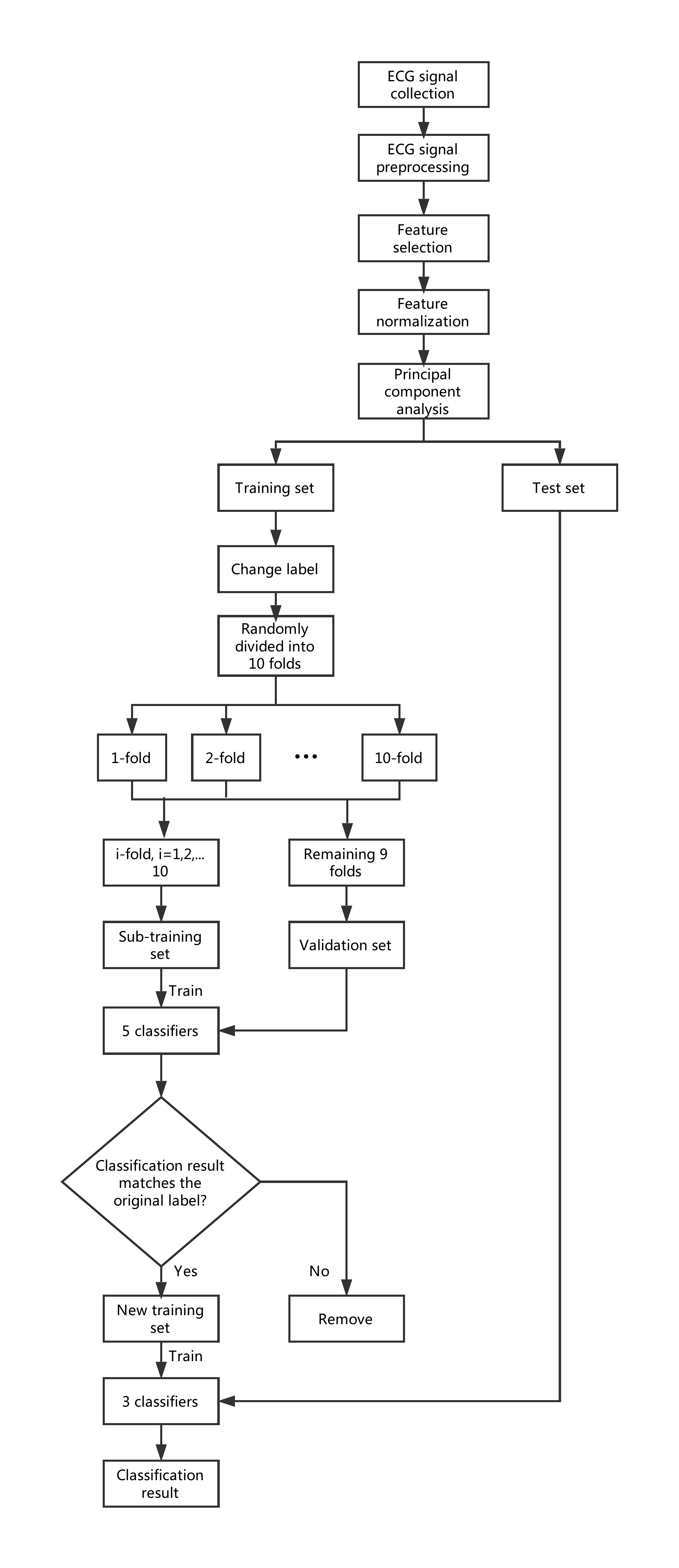}\\
  \caption{Flow chart of the proposed automatic training sample validation framework}
  \label{liuchengtu}
\end{figure}

As shown in Fig.~\ref{tuhua} and Fig.~\ref{liuchengtu}, with noise-free training set, we utilize the cooperative advantages of different classifiers to act as a filter for the training samples. The filter removes the mislabeled training samples and retains the correctly labeled ones with a 10-fold cross validation. Consequently, a new training set is fed to the final classifiers to acquire higher classification accuracy~\cite{Brodley:1999}. Then we create classifiers based on three machine learning algorithms, the naive bayes, the $k$ nearest neighbor and the linear discriminant analysis. These classifiers are used to verify the test set and calculate the classification accuracy. Afterwards, with the help of the cross validation, we dig out the mislabeled samples in the training set. The detailed procedures are the following.

{ All the training samples are randomly divided into 10 folds.} One fold serves as the validation set and the remaining 9 folds are the sub-training set for each time. Then, the sub-training set is fed to five machine learning algorithms~(SVM, C4.5, NB, KNN, LDA) to classify the validation set. The results of the classification are compared with the original label of the validation set samples. According to the comparison results, one can testify whether a certain sample in the validation set is a mislabeled sample or not. The above process is repeated 10 times. Each time the selected validation set is different so that all the samples in the training set can be verified. When all the mislabeled samples in the training set have been identified and removed, we get a new training set, retrain classifiers, and finally obtain satisfactory classification accuracies on the test set.

In the above procedure, different machine learning classifiers may give different validation results. So, we design three criterions to solve this problem

a.~Standard (1): If all five classifiers determine a single sample mislabeled, we regard it as truly mislabeled.

b.~~Standard (2): If four or more classifiers determine a single sample mislabeled, we regard it as truly mislabeled.

c.~Standard (3): If three or more classifiers determine a sample single mislabeled, we regard it as truly mislabeled.

Due to the contingency of a single experiment and its possible errors, in order to enhance the reliability of the experiment and improve the persuasiveness of this method, the experimental process should be repeated several times. The identification circumstances of mislabeled samples are shown detailedly in Tables 1-3.

 \begin{center}
 \scriptsize
 { Table~1\\ Detection performance of mislabeled samples with standard (1)}\\
\label{tab:1} \vskip 3pt
\newcommand{\rb}[1]{\raisebox{1.9ex}[-2pt]{#1}}
 \begin{tabular}{cccccc}
  \toprule
Noise~level & ANM & INM & AINM & $P_D$ & $P_{FA}$  \\
  \midrule
~~5\%  & 255  & 267 & 227 & 89\% & 16  \% \\
~~10\% & 510  & 463 & 424 & 83\% & 7.65\% \\
~~20\% & 1020 & 700 & 675 & 66\% & 2.45\% \\
~~30\% & 1530 & 829 & 796 & 52\% & 2.16\% \\
~~40\% & 2040 & 640 & 586 & 29\% & 2.65\% \\
  \bottomrule
 \end{tabular}
\end{center}

 \begin{center}
 \scriptsize
 { Table~2\\ Detection performance of mislabeled samples with standard (2)}\\
\label{tab:2} \vskip 3pt
\newcommand{\rb}[1]{\raisebox{1.9ex}[-2pt]{#1}}
 \begin{tabular}{cccccc}
  \toprule
Noise~level & ANM & INM & AINM & $P_D$ & $P_{FA}$  \\
  \midrule
~~5\%  & 255  &  334 &  239 & 93.73\% & 37\% \\
~~10\% & 510  &  572 &  474 & 93\% & 19.22\% \\
~~20\% & 1020 &  912 &  825 & 81\% & 8.53 \% \\
~~30\% & 1530 & 1344 & 1216 & 79\% & 8.37 \% \\
~~40\% & 2040 & 1435 & 1207 & 59\% & 11.18\% \\
  \bottomrule
 \end{tabular}
\end{center}

 \begin{center}
 \scriptsize
 { Table~3\\ Detection performance of mislabeled samples with standard (3)}\\
\label{tab:3} \vskip 3pt
\newcommand{\rb}[1]{\raisebox{1.9ex}[-2pt]{#1}}
 \begin{tabular}{cccccc}
  \toprule
Noise~level & ANM & INM & AINM & $P_D$ & $P_{FA}$  \\
  \midrule
~~5\%  & 255  &  873 &  251 & 98.43\% & 242\% \\
~~10\% & 510  & 1079 &  504 &  99\% & 112.75\% \\
~~20\% & 1020 & 1346 &  891 &  87\% &  44.61\% \\
~~30\% & 1530 & 2050 & 1401 &  92\% &  42.42\% \\
~~40\% & 2040 & 2567 & 1640 & 80\% &  45.44\% \\
  \bottomrule
 \end{tabular}
\end{center}

In Tables 1-3, ANM refers to the number of actual mislabeled samples, INM refers to the number of identified mislabeled samples, { AINM refers to the number of the mislabeled samples that are corrected identified.} $P_D$ refers to identification accuracy of mislabeled samples,
\begin{eqnarray}
P_D=\frac{AINM}{ANM}.
\end{eqnarray}
$P_{FA}$ refers to identification error rate,
\begin{eqnarray}
P_{FA}=\frac{INM-AINM}{ANM}.
\end{eqnarray}
We conclude that at the same noise level, $P_D$ obtains its lowest value in standard (1) and the highest value in standard (3). Standard (1) is a cautious standard, once one of the five classifiers determines a certain sample as correctly labeled, we regard it as the truly correctly labeled. Standard (3) is the most relaxed and gives more mislabeled samples. In contrast, $P_{FA}$ obtains its highest value in standard (1) and the lowest value in standard (3). Obviously, there is a trade-off between $P_D$ and $P_{FA}$, and the two objectives optimization is of particularly interest. In combination with these two aspects, standard (2) is the best standard. In Ref.\cite{Pasolli:2015}, when the label noise are 5\%, 10\%, 20\%, $P_D$ are 78.46\%, 78.40\%, 72.40\%, respectively, and $P_{FA}$ are 31.05\%, 15.65\%, 4.58\%, respectively.  In this paper, with standard (2), $P_D$ are 93.73\%, 93\%, 81\%, respectively, and $P_{FA}$ are 37\%, 19.22\%, 8.53\%, respectively. In summary, $P_D$ with the proposed method is significantly higher than Ref.\cite{Pasolli:2015}, with the cost of slightly higher $P_{FA}$.

Besides, with different standards in the procedure, we can identify the corresponding mislabeled samples and obtain new training sets. Meanwhile, new classifiers are updated to improve the classification accuracies. The details are shown in Tables 4-6:
\begin{center}
 \scriptsize
 { Table~4\\ Classification accuracy { obtained} by naive bayes classifier on the ECG beats for all the considered training set scenarios}\\
\label{tab:4} \vskip 3pt
\newcommand{\rb}[1]{\raisebox{1.9ex}[-2pt]{#1}}
 \begin{tabular}{cccccc}
  \toprule
  &\multicolumn{5}{c}{Naive~bayes}\\
\cmidrule(l){2-6} \rb{~~Noise~level} & NF & IF & S1 & S2 & S3 \\

\midrule
~~0    & 73.30\% &   --    &   --    &   --    &   --    \\
~~5\%  & 71.96\% & 73.13\% & 74.67\% & 75.09\% & 75.48\% \\
~~10\% & 69.33\% & 73.20\% & 74.16\% & 75.53\% & 75.23\% \\
~~20\% & 62.01\% & 73.04\% & 71.60\% & 74.15\% & 73.87\% \\
~~30\% & 51.85\% & 73.06\% & 64.95\% & 72.34\% & 70.71\% \\
~~40\% & 38.77\% & 72.74\% & 49.68\% & 57.87\% & 54.79\% \\
  \bottomrule
 \end{tabular}
\end{center}

\begin{center}
 \scriptsize
 { Table~5\\ Classification accuracy { {obtained}} by KNN classifier on the ECG beats for all the considered training set scenarios}\\
\label{tab:5} \vskip 3pt
\newcommand{\rb}[1]{\raisebox{1.9ex}[-2pt]{#1}}
 \begin{tabular}{cccccc}
  \toprule
  &\multicolumn{5}{c}{KNN}\\
\cmidrule(l){2-6} \rb{~~Noise~level} & NF & IF & S1 & S2 & S3 \\

\midrule
~~0    & 97.50\% &   --    &   --    &   --    &   --    \\
~~5\%  & 96.60\% & 97.29\% & 97.00\% & 97.25\% & 95.22\% \\
~~10\% & 94.97\% & 97.20\% & 96.91\% & 96.80\% & 94.38\% \\
~~20\% & 87.84\% & 97.12\% & 93.97\% & 95.53\% & 94.73\% \\
~~30\% & 77.48\% & 96.96\% & 86.69\% & 92.22\% & 88.81\% \\
~~40\% & 64.16\% & 96.86\% & 71.20\% & 78.15\% & 71.12\% \\
  \bottomrule
 \end{tabular}
\end{center}

\begin{center}
 \scriptsize
 { Table~6\\ Classification accuracy { obtained} by LDA classifier on the ECG beats for all the considered training set scenarios}\\
\label{tab:6} \vskip 3pt
\newcommand{\rb}[1]{\raisebox{1.9ex}[-2pt]{#1}}
 \begin{tabular}{cccccc}
  \toprule
  &\multicolumn{5}{c}{LDA}\\
\cmidrule(l){2-6} \rb{~~Noise~level} & NF & IF & S1 & S2 & S3 \\

\midrule
~~0    & 74.50\% &   --    &   --    &   --    &   --    \\
~~5\%  & 73.88\% & 74.27\% & 74.28\% & 74.38\% & 73.93\% \\
~~10\% & 73.41\% & 74.35\% & 74.12\% & 74.23\% & 73.35\% \\
~~20\% & 71.27\% & 74.62\% & 73.27\% & 73.82\% & 72.49\% \\
~~30\% & 63.78\% & 74.35\% & 68.87\% & 70.95\% & 67.82\% \\
~~40\% & 48.89\% & 74.40\% & 55.16\% & 58.27\% & 50.12\% \\
  \bottomrule
 \end{tabular}
\end{center}

\begin{figure}
  \centering
  \includegraphics[width=6cm]{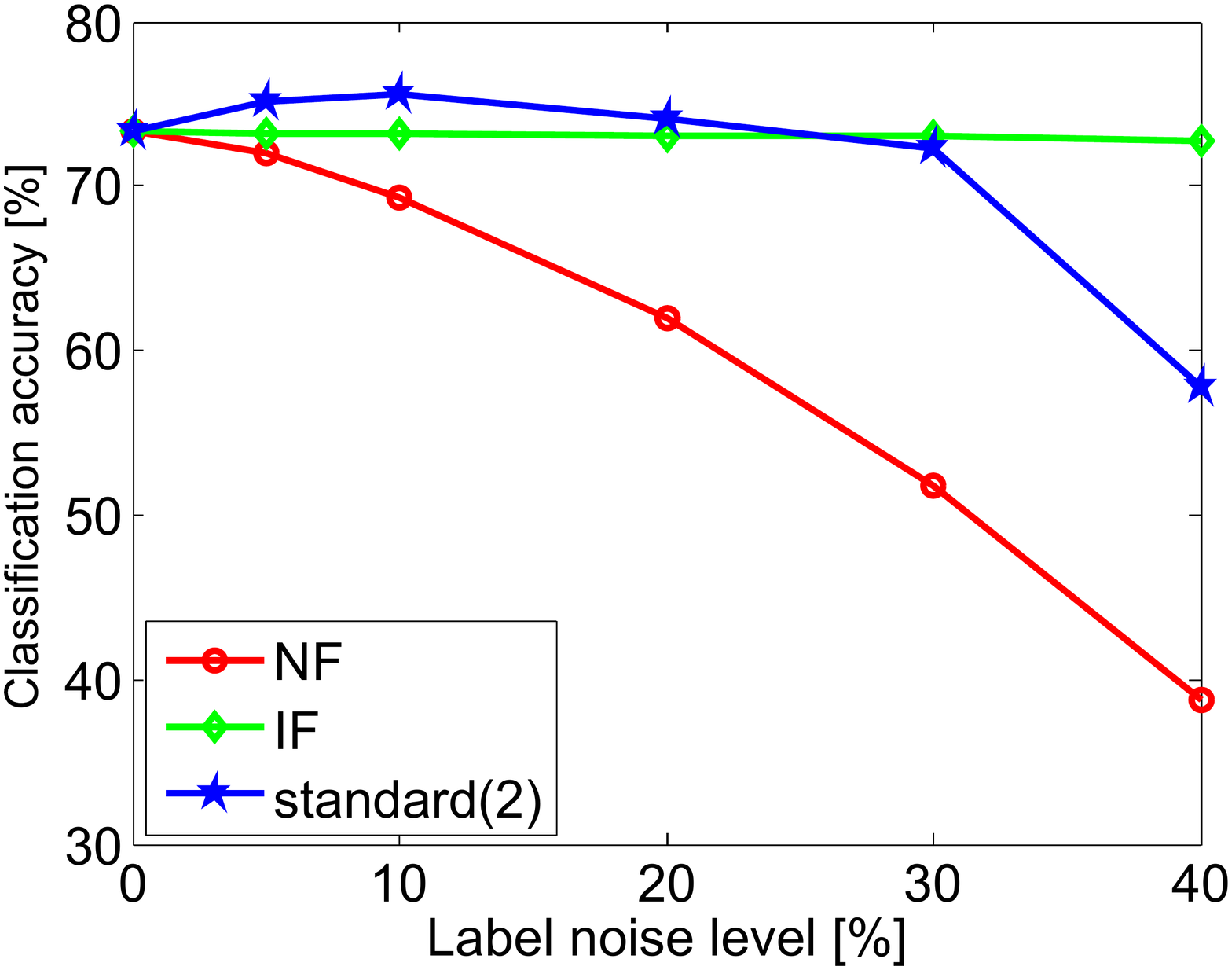}\\
  \caption{NB classification accuracy}
  \label{NB}
\end{figure}

\begin{figure}
  \centering
  \includegraphics[width=6cm]{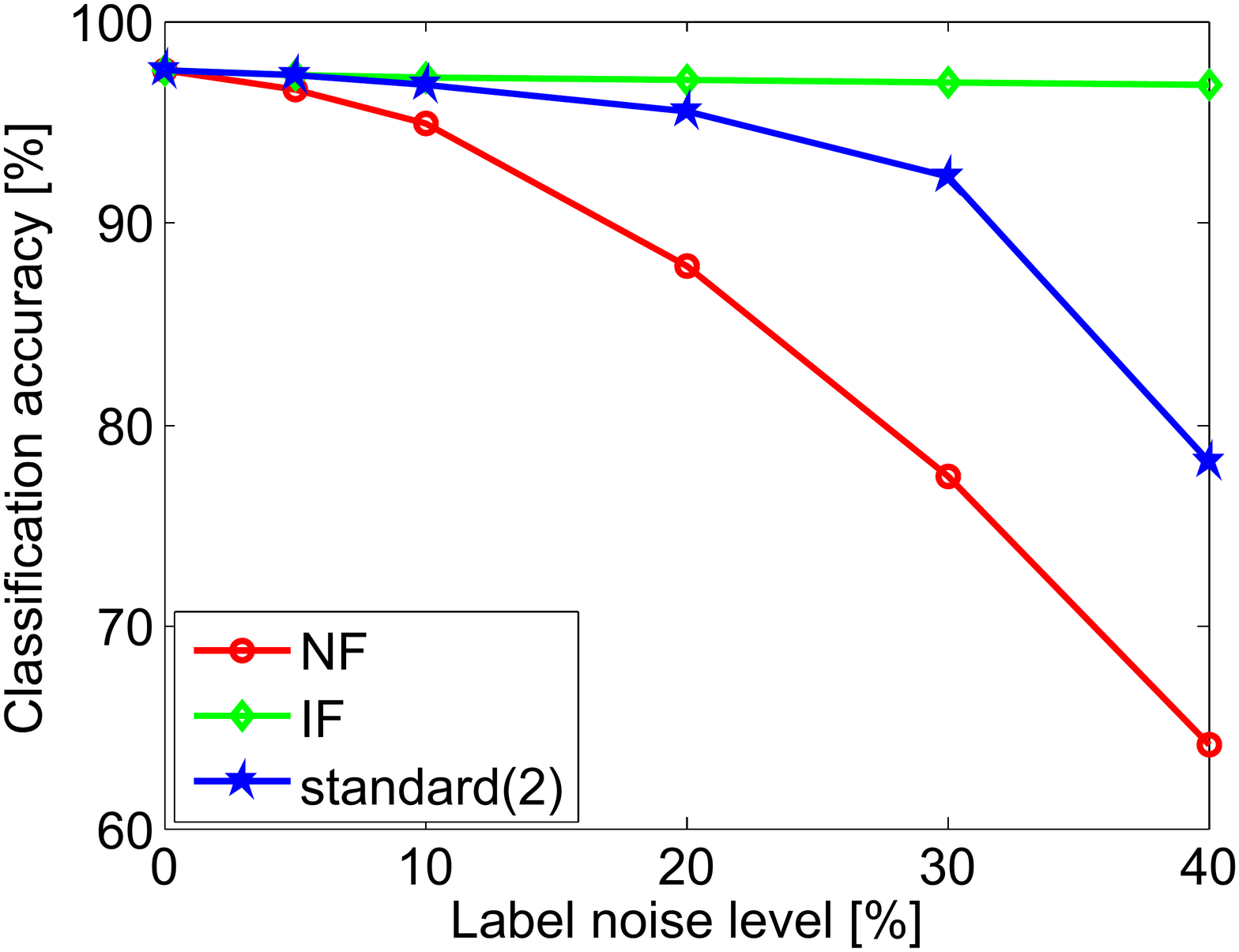}\\
  \caption{KNN classification accuracy}
  \label{KNN}
\end{figure}

\begin{figure}
  \centering
  \includegraphics[width=6cm]{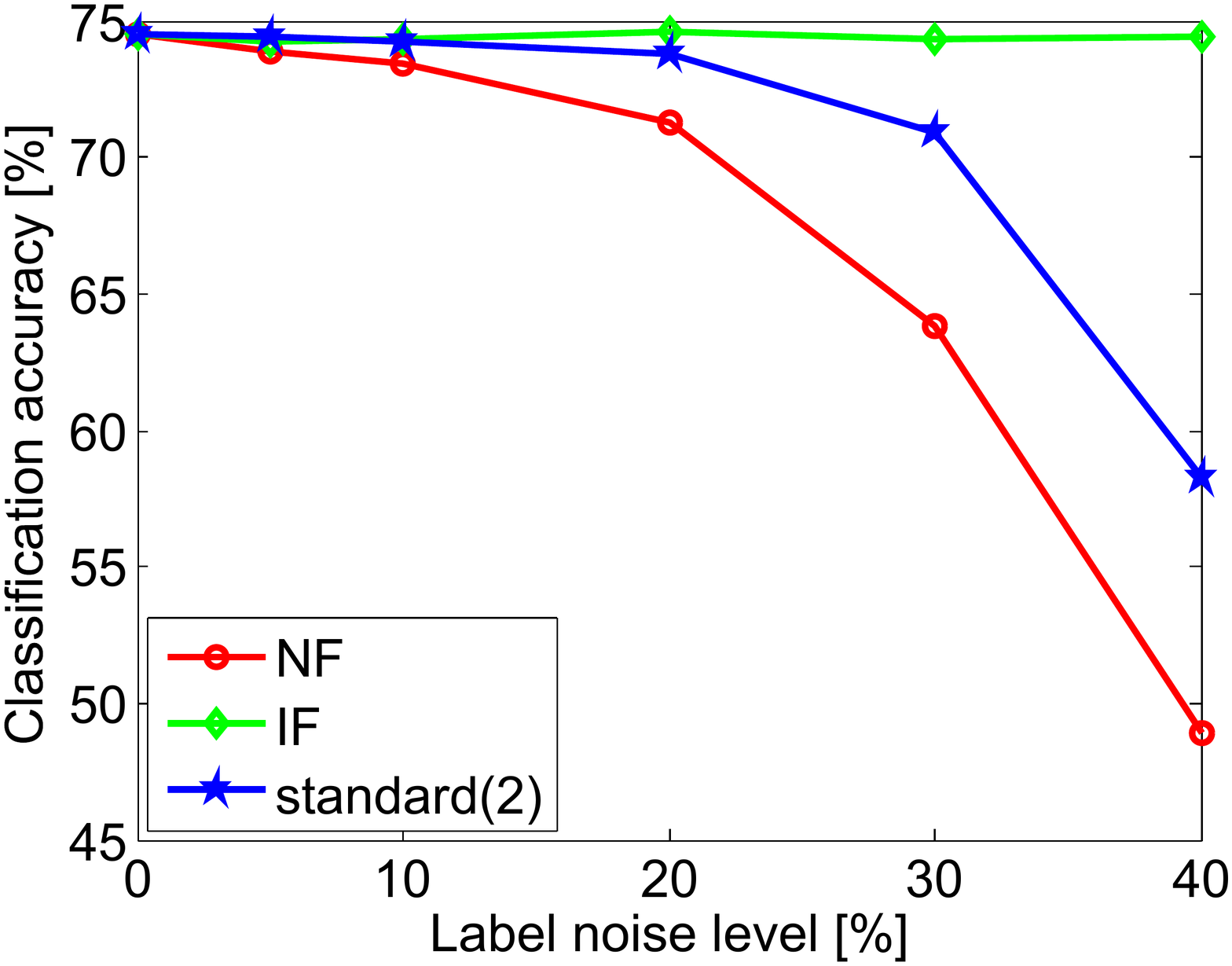}\\
  \caption{LDA classification accuracy}
  \label{LDA}
\end{figure}

In Tables 4--6 and Figs.~\ref{NB,KNN,LDA} , NF refers to no filtering. IF means ideal filtering~(artificially removes all the mislabeled samples from the training set). S1, S2, S3 represent standard~(1), standard~(2) and standard~(3), respectively. Obviously, the classification accuracies of NB, KNN and LDA reduce remarkably with the increase of the label noise level in the training set without filtering. In the case of ideal filtering , the classification accuracies are improved to almost the same level as the noise level equals 0. With different standards, the proposed method gets different filtering results. Specifically, the classification accuracies on the test set have been significantly improved. In some individual cases, such as the label noise equals 10\%, the accuracy of the proposed method is even higher than that of the ideal filtering, because our method can not only filters out the artificially added label noise, but also removes some undiscovered noise. These noises may come from feature extraction or other processes after ECG denoising. From the experimental results in Tables 4--6, we note that standard~(2) has the best classification accuracy on the test set, which is consistent with what we mentioned in Sec.IV(B). As shown in Figs. 6--8, if the mislabeled samples existed in the training set is lower than 20\%, the classification accuracy can be increased to the same level as there is no mislabeled sample in the training set with the help of the proposed method. If the proportion is 30\%, the classification accuracy is slightly lower than that with no mislabeled sample. If the proportion meets 40\%, the classification accuracy is still much higher than the circumstances without filtering.


\section{Conclusion}
In this paper, we investigate how to identify and eliminate the mislabeled training samples that are widely existed in ECG analysis. We use the cross validation method to improve the identification rate and the classification accuracy of ECG signal. Simulation results demonstrate the effectiveness of our proposed method. The mislabeled samples in the training set directly damage classifiers and their classification accuracy seriously. We use five different machine learning classifiers to classify the validation set, and to improve the reliability of the mislabeled samples identification. Especially, if the label noise level is not higher than 20\%, the classification accuracy can be improved to the same level as there is no mislabeled samples in the training set. It is noted that, the mislabeled noises are randomly added rather from the similarity of different arrhythmias in the actual ECG waveform.
There are also some points needing for improvement in our further works. For example, the computational load is relatively expensive due to the large number of classifiers, and also, our method can not be competent for too high noise levels. Further researches are necessary to address these problems and to make this method more practical and effective.

\vskip 12pt
 {\fontsize{7.8pt}{9.4pt}\selectfont
}

\end{document}